\documentclass{aa}
\usepackage{graphicx}
\newcommand\simlt{\lower.5ex\hbox{$\; \buildrel < \over \sim \;$}}
\newcommand\simgt{\lower.5ex\hbox{$\; \buildrel > \over \sim \;$}}
\newcommand\halpha{H$\alpha$}
\newcommand\rs[1]{_\mathrm{#1}}
\newcommand\zr{\rs{o}}  \newcommand\st{\rs{*}}

\begin{document}

\title{Pulsar bow-shock nebulae}
\subtitle{III. Inclusion of a neutral component, and \halpha \ luminosity.}
\author{N. Bucciantini} % \inst{1} 

\offprints{N. Bucciantini; \\
\email{niccolo@arcetri.astro.it}}
\institute{Dipartimento di Astronomia e Scienza dello Spazio,
             Universit\`a degli Studi di Firenze, \\
             Largo E. Fermi 2, I-50125 Firenze, Italy}

\date{Received 5 June 2002 / Accepted 28 June 2002}
\abstract{The interaction of the wind from a pulsar (or more generally from a star) with the ambient medium gives rise to the formation of a bow-shock nebula. We present a model of adiabatic bow-shock nebulae, including the presence of a neutral component in the interstellar medium. As demonstrated in a previous paper (Bucciantini \& Bandiera \cite{bucciantini01}, hereafter Paper I), the velocity and the luminosity of the pulsar, as well as the characteristics of the interstellar medium (ISM), play an essential role in determining the properties of bow shock nebulae. In particular, the Balmer emission is strongly affected by the rates of hydrogen charge-exchange and collisional excitation. So far, only one-component models have been proposed, treating the neutrals as a small perturbation. The distribution of neutral hydrogen in the nebula, derived by our model, is then used to determine how the \halpha \ luminosity varies along the bow shock. We find that the luminosity trend is different for charge-exchange and collisional excitation, and that processes like diffusion or ionization may reduce the emission in the head of the nebula.
}
\maketitle
%---------------------------------------------------------------------------
\section{Introduction}
Pulsars are known to be sources of ultrarelativistic magnetized winds (see e.g. Michel \& Li \cite{michel99}). If during the supernova explosion the newly formed neutron star acquires a high velocity, it may escape from the supernova remnant before its pulsar activity has ceased. In such a case the pulsar wind interacts with the interstellar medium (thereafter ISM), possibly forming a steady flow, with a bow shock in the direction of the stellar motion.
So far only four such nebulae are known: PSR 1957+20 (Kulkarni \& Hester \cite{kulkarni88}), PSR 2224+65 (Cordes et al. \cite{cordes93}, Chatterjee \& Cordes \cite{chatterjee02}), PSR J0437+4715 (Bell \cite{bell95}), and PSR 0740-28 (Jones et al. \cite{jones01}). Probably even the nebula associated with the neutron star RX J185635-3754 (van Kerkwijk \& Kulkarni \cite{vankerkwijk01}) may be included in the list, even if there is no agreement whether the nebula is a bow shock or the result of photoionization.

All the nebulae discovered have been detected in Balmer lines (mostly \halpha ). A pure Balmer spectrum is the signature of a non-radiative shock moving through a partially neutral medium (Chevalier \& Raymond \cite{chevalier80}, Chatterjee \& Cordes \cite{chatterjee02}). When recombination times are long the relevant processes for optical emission are exciting charge-exchange and collisional excitation of neutral H atoms.
 
As mentioned in a previous paper of this series (Bucciantini \cite{bucciantini02}, hereafter Paper II), the ISM is seen by the pulsar as a plane parallel flow with a constant density (at least on the length scale of the bow shock). If the ISM is partially neutral, and if the neutral atoms have mean free paths comparable with, or longer than the size of the bow shock a pure fluid treatment is invalid. The interaction of the relativistic magnetized pulsar wind with the ionized component of the ISM is mediated by the magnetic field advected by the pulsar wind and compressed in the head of the nebula. The neutral atoms interact with the ions (mainly protons) thermalized in the bow shock. The penetration thickness for the H atoms changes from pulsar to pulsar depending on their velocity, their luminosity as well as the density and the neutral fraction of the ISM; and the same holds for the diffusion and ionization (Paper I). Thus to have a more realistic picture of bow-shock nebulae in this regime, it is necessary to include the effect of mass loading from a neutral component as well as the diffusion and ionization of dragged atoms.

As a starting point we have used a two thin layers  bow shock model, for which an analytic solution may be derived (Comeron \& Kaper \cite{comeron98}). In Paper II we have verified, using a numerical hydrodynamic code, that this analytic model gives reasonable estimates for the values of surface density and tangential velocity in the external layer. These estimates describe the fluid structure in the external layer and have been used to model the interaction and the evolution of the neutral atoms.

Once the density of the neutral H atoms in the external layer has been derived, it is used to evaluate the \halpha \ luminosity of the nebula, and its variation along the bow-shock. The comparison of the observed luminosity with that obtained from a thin layer model may be used to determine the properties of the local ISM, as well as those of the pulsar wind.
In Section 2 we present the equations we have used to derive the distribution of neutral atoms, and the solutions derived integrating these equations; in Section 3 we derive the \halpha \ luminosity of the nebula and discuss what informations may be obtained from observations. 

\section{The model}
\subsection{The zeroth order two thin layer solution, and the effect of dragging charge-exchange}

As said in Paper I, the parameters defining the scenario of a runaway pulsar moving supersonically in the ISM are: the pulsar energy losses, $L$; its velocity in the rest frame of the ambient medium, $v\zr$; the density respectively of the ionized and neutral component of the ambient medium, $n\zr$, $\Xi\zr n\zr$. A supersonic motion implies that the ram pressure of the ISM is greater than its thermal pressure, so that the latter may be neglected in the model. For typical pulsar velocities the related Mach number $M\zr$ is $\simgt 4$ (for a diffuse warm ionized medium).

In Paper II we stressed the fact that a simple bow-shock model (Wilkin \cite{wilkin96}), may not be adequate for the case of a pulsar. The pulsar wind is magnetized; thus the basic assumption of the Wilkin solution (efficient mixing between stellar wind and ambient medium) is not satisfied. For a better description we need to follow separately the two layers, one containing the shocked stellar wind and the other the shocked external medium, allowing two different tangential velocities. In this work we are mainly interested in the behaviour of the external layer of thermalized ISM.
%%%%%%%%%%%%
\begin{figure}
\centering{

    \resizebox{\hsize}{!}{\includegraphics{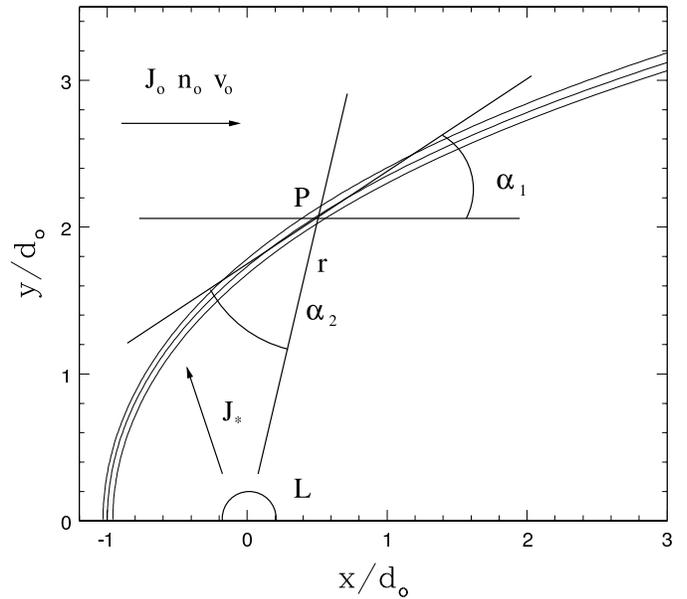}}}
   \caption{Geometry of the two thin layer model (completely passing case). $J\zr$ and $n\zr$ are the momentum flux, and the numerical density of the ambient medium; $J_{*}$ and $L$ are the stellar wind momentum flux and the pulsar luminosity; $v\zr$ is the pulsar velocity with respect to the external medium. }
   \label{fig:1}
\end{figure}
%%%%%%%%%%%%

If we retain the thin layer assumption, equations describing the shape, tangential velocity and surface density, can be derived from the momentum and mass conservation. We shall start from the freely crossing case (the neutrals do not interact), then adding on it the effect of mass loading from a neutral component. In fact in typical cases the density of neutral hydrogen may be of the same order  as that of protons.

Let us first summarize the freely crossing case. With reference to Fig.~\ref{fig:1}, let us consider a point (P) on the layer with distance $r$ from the star, where the incident fluxes of momentum are $J\zr$ and $J\st$. The conservation of perpendicular momentum gives:
\begin{equation}
  \label{eq:mompep}
J\zr\sin^{2}\alpha_{1}-J\st\sin^{2}\alpha_{2}=\frac{\sigma_{fi}v_{f}^{2}+\Phi_{r}}{R},
\end{equation}
where $\sigma_{fi}$ is the surface density of the ionized component in the external layer, $v_{f}$ is the tangential velocity in the external layer, $R$ is the curvature radius of the front, and $\Phi_{r}$ is the momentum flux in the internal layer. For the tangential component of the fluxes in the external and internal layer we have respectively:
\begin{equation}
  \label{eq:momtex}
J\zr\sin\alpha_{1}\cos\alpha_{1}=\frac{d(\sigma_{fi}v_{f}^{2})}{ds},
\end{equation}
\begin{equation}
  \label{eq:momtin}
J\st\sin\alpha_{2}\cos\alpha_{2}=\frac{d\Phi_{r}}{ds},
\end{equation}
where $ds$ is the line element. Finally the mass conservation equation for the external layer, where surface density is required for the evaluation of penetration thickness, reads:
\begin{equation}
  \label{eq:massc}
\sigma_{fi}v_{f}= \frac{J\zr}{2v\zr}y.
\end{equation}
These equations may be written in a simpler form using dimensionless variables. The lengths $y$ and $x$ scale with the distance to the stagnation point (Cordes \cite{cordes96}):
\begin{equation}
  \label{eq:stpoint}
d\zr=\sqrt{J\st/J\zr}=\sqrt{L/4\pi cn\zr v\zr^{2}},
\end{equation}
while the velocity and surface density in the external layer are respectively scaled with $v\zr$ and $n\zr\, d\zr$, and the momentum flux in the internal layer with $J\zr\, d\zr$. Equations \ref{eq:mompep} to \ref{eq:massc} may be written:
\begin{eqnarray}
(\sigma_{fi}v_{f}^{2}+\Phi_{r})\frac{dz}{dy}-n+\frac{n(yz-x)^{2}}{r^{4}}=0,\nonumber\\
\label{eqsis}
\frac{d(y\sigma_{fi}v_{f}^{2})}{y\,dy}-\frac{z}{n}=0,\\
\frac{d(y\Phi_{r})}{y\,dy}-\frac{(xz+y)(yz-x)}{n r^{4}}=0,\nonumber\\
2\sigma_{fi}v_{f}-y=0.\nonumber
\end{eqnarray}
Where $z=dx/dy$ and $n=(1+z^{2})^{-1/2}$.
This set of equations can be integrated numerically, starting from the stagnation point, and using a Taylor expansion near the apex. It is remarkable to see that these equations do not contain any free parameter, so there is only one solution in these dimensionless coordinates (solid lines of Figs.~2 and 3).

Once the freely crossing solution has been obtained from the integration of Eqs.~\ref{eqsis}, it is possible to introduce the effects of a neutral component of ISM (that we assume to be mainly composed by neutral hydrogen). As said in Paper I, for typical pulsar velocities the main process responsible for the dragging of neutral H in the external layer is charge-exchange. Consequences of the presence of a neutral component are: 1- a new source of perpendicular and tangential momentum, 2- the need for a further mass conservation equation for the neutrals. We assume that, once a neutral atom has been captured in the external layer, it can neither escape via diffusion nor be ionized (an assumption whose implications will be analyzed in section 2.2), and that there is no shear between the neutral and ionized material.  The penetration thickness of the external layer is not constant but tends to increase moving away from the stagnation point (Paper II). The fraction of neutrals stopped in this layer is:
\begin{equation}
 F=1-{\rm exp}(-\sigma_{cx}(v_{rel})\sigma_{fi}v_{rel}/v\zr\sin{\alpha_{1}}),
\end{equation}
where $\sigma_{cx}$ is the cross section of charge-exchange, $v_{rel}$ is the relative velocity between the incoming H atoms and the hot shocked plasma and we have assumed a constant state (i.e. no variations of density, temperature, velocity) across the external layer. We want to stress that the constant state assumption is valid only near the head of the bow shock where the geometrical thickness of the layer is small compared to the size of the nebula. In order to evaluate the relative velocity we assume that all the ionized material in the external layer moves tangentially with a speed $v_{f}$: this assumption is good (Paper I) if the penetration thickness is not too high, otherwise it would be better to use the post shock velocity (according to the Rankine-Hougoniot jump conditions). However the use of the post shock velocity is justified only in case $F \simeq1$, anyway giving results not too different from the other case.  
The set of Eqs.~\ref{eqsis}, becomes:
\begin{eqnarray}{}
((\sigma_{fi}+\sigma_{fn})v_{f}^{2}+\Phi_{r})\frac{dz}{dy}-n(1+\Xi\zr  F)+\frac{n(yz-x)^{2}}{r^{4}} & = & 0,\nonumber\\
\label{eqsis2}
\frac{d(y(\sigma_{fi}+\sigma_{fn})v_{f}^{2})}{y\,dy}-\frac{z}{n}(1+\Xi\zr F)=0 ,&\\
\frac{d(y\Phi_{r})}{y\,dy}-\frac{(xz+y)(yz-x)}{n r^{4}}=0,\nonumber\\
\frac{d(y\sigma_{fi}v_{f})}{dy}-y=0,\nonumber\\
\frac{d(y\sigma_{fn}v_{f})}{dy}-y\Xi\zr F=0,\nonumber
\end{eqnarray}
where $\sigma_{fn}$ is the surface density of the neutral component in the external layer. It is convenient to parametrize $F$ using a power law for the cross section in the typical range of pulsar velocities (Newman \& al. \cite{newman82}, McClure \cite{mcclure66}, Peek \cite{peek66}):
\begin{equation}
\label{eqcx}
\sigma_{cx}(v_{rel})=\sigma_{cx}(v_{rel,0})\left(\frac{v_{rel}}{v_{rel,0}}\right)^{-0.37},
\end{equation}
(where we have taken $v_{rel,0}=1$) so that:
\begin{equation}
 F=1-F\zr^{(v_{rel}(y)^{0.63}\sigma_{fi}n/1.25)}.
\end{equation}
Therefor $1-F\zr$ corresponds to the fraction of neutrals captured at the stagnation point, in the fully ionized case, which defines the properties of the true physical nebula (the factor $1.25$ corresponds to the surface density of the external layer on the axis, in the freely crossing case). 

Contrary to the freely crossing case, Eqs.~\ref{eqsis2} cannot be integrated from the head, because they are not closed (i.e. in the Taylor expansion  of the equations the terms of order $\cal{O}$($y^{n}$) are function of those of order $\cal{O}$($y^{n+1}$)). Instead we have used a perturbative method: from the freely crossing solution, we evaluate a first approximation of $F$; this approximation is then used in Eqs.~\ref{eqsis2} instead of the self-consistent expression, both to obtain the Taylor expansion and to integrate numerically the equations;  we repeat iteratively till $F$ converges.

In Fig.~2 we present the solution of Eqs.~\ref{eqsis2} for the case $\Xi\zr=1$ (an intermediate case between the highly ionized and highly neutral ISM) for different penetration thicknesses in the head. The solid line represents the freely crossing case (it is a good approximation if the thickness remains $\ll 1$ all along the nebula), while the dotted line refers to the completely dragging case (which is simply a bow shock scaled for a greater external density). The stagnation point distance from the star decreases with increasing values of $F\zr$, and its value is:
\begin{equation}
d\simeq d\zr/\sqrt{(1+\Xi\zr(1 - F\zr))}.
\end{equation}
The self-consistency level of our solution decreases moving to the tail of the nebula, away from the star, because the thin layer approximation is no longer acceptable, an so the constant (perpendicularly to the bow-shock) internal state assumption lose its validity.

Fig.~3 shows the surface density of the ionized component in the$\Xi\zr=1$ case: also here the solid line represents the freely crossing case while the dotted line is the completely dragging solution. It is evident that the solution with highest $F$ tends to the dotted line in the tail of the nebula.
Fig.~4 gives the surface density of the neutral component in the various cases obtained from Eqs.~\ref{eqsis2}. Once the neutral and the ionized component density has been obtained it is quite easy to derive the \halpha \ luminosity along the nebula. However, before doing this it is necessary to consider the effect of other processes like diffusion and ionization (see Paper I) which modify the neutral distribution. 
%%%%%%%%%%%%
\begin{figure}
\centering{

    \resizebox{\hsize}{!}{\includegraphics{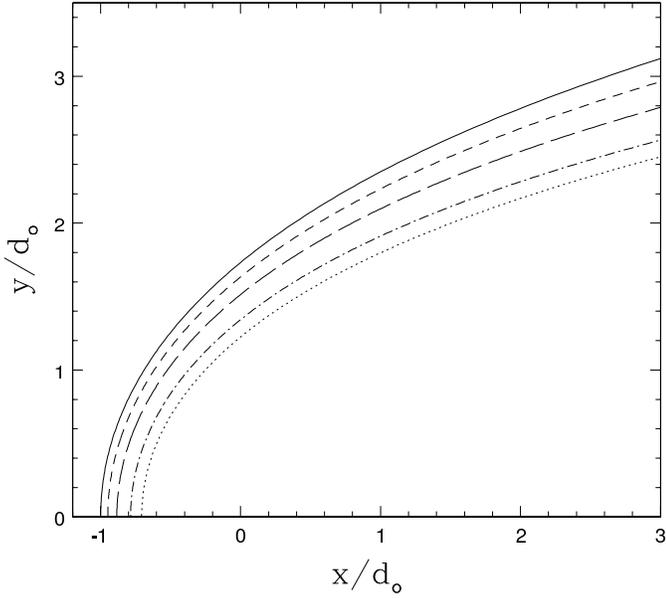}}}
   \caption{Shape of the bow shock for various penetration thickness in the head. The solid line is the freely crossing case; the dotted one is the completely dragging case. The short-dashed, long-dashed and dot-dashed lines refer respectively to three values (0.1, 0.3, 1.) of the parameter $\sigma_{cx}(v_{rel,0})n\zr d\zr$.}
   \label{fig:2}
\end{figure}
%%%%%%%%%%%%
%%%%%%%%%%%%
\begin{figure}
\centering{

    \resizebox{\hsize}{!}{\includegraphics{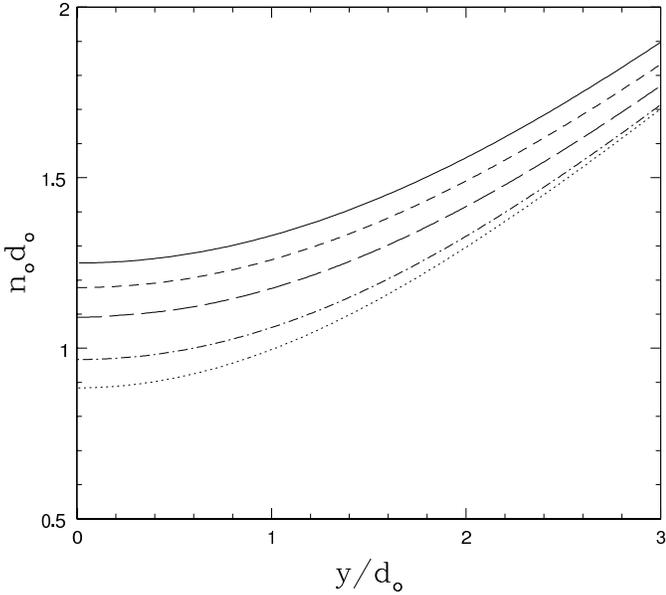}}}
   \caption{Surface density of the ionized component in the external layer the bow shock for various penetration thickness in the head ($\Xi\zr=1$). The various lines refer to the cases of Fig.~2. }
   \label{fig:3}
\end{figure}
%%%%%%%%%%%%
%%%%%%%%%%%%
\begin{figure}
\centering{

    \resizebox{\hsize}{!}{\includegraphics{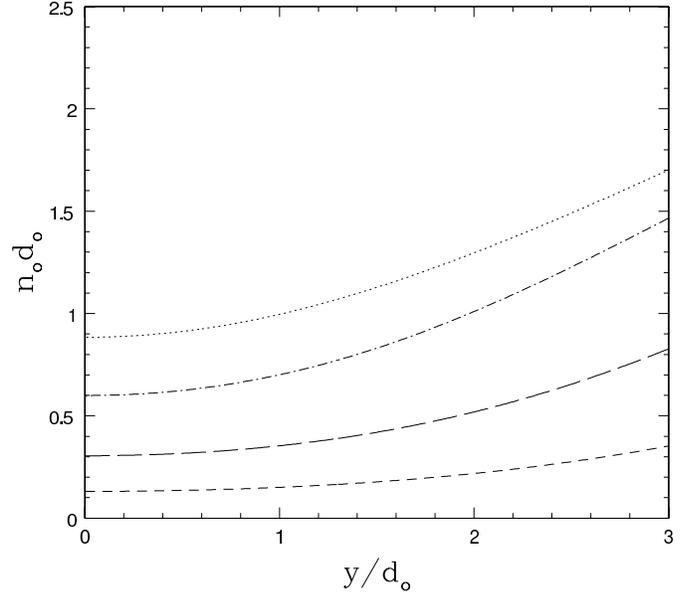}}}
   \caption{Surface density of the neutral component in the external layer the  the bow shock for various values of the penetration thickness in the head ($\Xi\zr=1$). The various lines refer to the cases of Fig.~2.}
   \label{fig:4}
\end{figure}
%%%%%%%%%%%%

\subsection{Importance of diffusion and ionization and their inclusion in the model}

As previously mentioned the neutral H atoms, once stopped by charge-exchange with thermalized protons of the external layer, may diffuse or be ionized as a consequence of processes of charge-exchange and ionization in the internal hot plasma. Before introducing these effects in the equations it is necessary to discuss the physical regimes of the pulsar bow shock nebula, and their variations along the nebula. Comparing the time scale of the flow $\tau_{flux}\sim d\zr/v\zr$ with, respectively, the time scale of diffusion and ionization, tells what are the importance of these two processes in the various pulsar cases.

The time scale for diffusion is:
\begin{equation}
\tau_{diff}\sim \frac{\Delta\;\sigma_{fi}\sigma_{cx}(v_{th})}{v_{th}},
\end{equation}
where $\Delta$ is the geometrical thickness of the external layer, the cross section of diffusion $\sigma_{cx}$ is that of charge-exchange, and the thermal velocity $v_{th}$ is that obtained from the Rankine-Hugoniot condition on the bow shock $\sim0.8\,v\zr\sin{\alpha_{1}}$. Moving away from the head this time scale tends to increase, so that if it is greater than the flow time on the head, diffusion may not be taken into account.
The time scale for ionization is:
\begin{equation}
\tau_{ion}\sim \frac{\Delta}{\sigma_{fi}\sigma_{ion}(v_{th})v_{th}},
\end{equation}
where $\sigma_{ion}$ is the cross section of ionization, that, in the range of pulsar velocities, can be approximated by the relation (Ptak \& Stoner \cite{ptak73}):
\begin{equation}
\label{eqion}
\sigma_{ion}(v_{th})=\sigma_{ion}(v_{th,0})\left(\frac{v_{th}}{v_{th,0}}\right)^{2.9}.
\end{equation}
Also this time scale increases in the tail of the nebula, mainly as a consequence of the decreasing of the thermal velocity.

In the case of very luminous pulsars it is also necessary to determine the relative importance of diffusion and ionization, because atoms are ionized before the may diffuse (Paper I). The ratio $\tau_{diff}/\tau_{ion}$ tends to decrease away from the head because the cross section of ionization has a steeper  velocity dependence than that of charge-exchange.

The effect of diffusion and ionization may be included in the equations to give a more realistic model, but some simplifications are necessary. We have assumed that atoms in the external layer have the same probability to diffuse in the internal stellar wind region as to diffuse in the external region of unperturbed ISM. In reality there may be an asymmetry in the diffusion not taken into account in our model, however the results of its inclusion are not significatively different, both in terms of surface densities and shape of the nebula. Diffusion acts in two ways: 1- reducing the surface density of neutral atoms, 2- bringing away from the layer an amount of tangential momentum, proportional to the number of diffused atoms. To treat properly diffusion we need to know the distribution of neutrals in the layer, but thin-layer models lack this piece of information. Contrary to diffusion, ionization does not change the total surface density in the external layer, but simply changes neutrals into ions. This may modify the geometry of the nebula because the increased ionized surface density makes the layer thicker to the penetration of neutrals. Using Eqs.~\ref{eqcx} and \ref{eqion} for the parametrization of the cross sections  we find that the rate of diffusion $C$ and ionization $B$ may be written:
\begin{eqnarray}{}
 C & = &  C\zr\frac{\sigma_{fn}\Delta\zr}{n^{0.37}\sigma_{fi}\Delta},\\
 B & = & B\zr\frac{\sigma_{fn}\sigma_{fi}\sin{\alpha_{1}}^{3.9}\Delta\zr}{\Delta},
\end{eqnarray}
where all the unknown parameters are enclosed in the quantities  $C\zr$ and $B\zr$, and $\Delta\zr$ is the thickness of the layer along the axis in the freely crossing case. From a numerical simulation (Paper II) we find the parametrization $\Delta=\Delta\zr d(1+y^{2}/5)/d\zr$.
Eqs. \ref{eqsis2} then become:
\begin{eqnarray}{}
 ((\sigma_{fi}+\sigma_{fn})v_{f}^{2}+\Phi_{r})\frac{dz}{dy} -n(1+\Xi\zr  F)+ \frac{n(yz-x)^{2}}{r^{4}} & = & 0 ,\nonumber\\
\label{eqsis3}
 \frac{d(y(\sigma_{fi}+\sigma_{fn})v_{f}^{2})}{y\,dy} -\frac{z}{n}(1+\Xi\zr\  F)+ C v_{f}  =  0 ,  &\\
\frac{d(y\Phi_{r})}{y\,dy}-\frac{(xz+y)(yz-x)}{n r^{4}}=0,\nonumber\\
\frac{d(y\sigma_{fi}v_{f})}{dy}-y(1 + B)=0,\nonumber\\
\frac{d(y\sigma_{fn}v_{f})}{dy}-y( \Xi\zr\ F - C - B)=0.\nonumber
\end{eqnarray}
They can be integrated in the same way as discussed in the previous subsection, in fact the quantities $C$ and $B$ may be evaluated self-consistently.

Figs. 5 and 6 show respectively the effect of diffusion and of ionization on the surface density of the neutral component in the external layer, for various values of the coefficient $C\zr$ and $B\zr$. Both diffusion and ionization tend to reduce the density of H atoms. However this effect is reduced in the tail: diffusion decreases both as a consequence of the greater thickness of the layer and as a conseguence of the reduction of the mean free path of neutral H atoms; ionization simply vanishes because the cross section decreases rapidly with the thermal velocity. Fig.~7 shows the surface density of the ionized component: in case of hight ionization rates it tend to decrease in the head with respect to the values on the axis. This influences the \halpha \ luminosity, making it less that what may be derived from a simple case with only dragging charge-exchange.

 %%%%%%%%%%%%
\begin{figure}
\centering{

    \resizebox{\hsize}{!}{\includegraphics{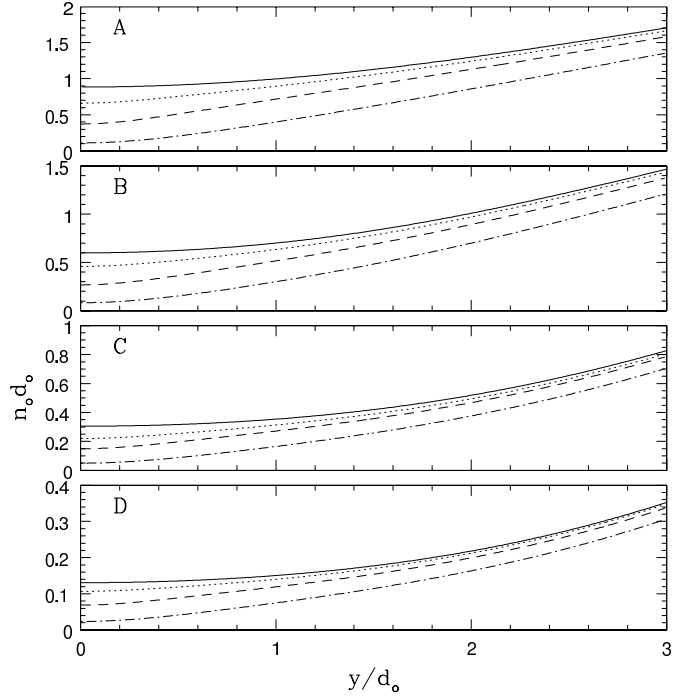}}}
   \caption{The effect of diffusion on the surface density for the neutral component ($Xi\zr=1$). Panels refer to: A to completely dragging case, B C and D respectively to the values 1, 0.3, 0.1 of the parameter $\sigma_{cx}(v_{rel,0})n\zr d\zr$. Continuous lines are the solutions without diffusion, dotted lines have $ C\zr = 0.25$, dashed lines have $C\zr = 0.8$, dash-dotted lines have $ C\zr = 4$.}
   \label{fig:5}
\end{figure}
%%%%%%%%%%%%
 %%%%%%%%%%%%
\begin{figure}
\centering{

    \resizebox{\hsize}{!}{\includegraphics{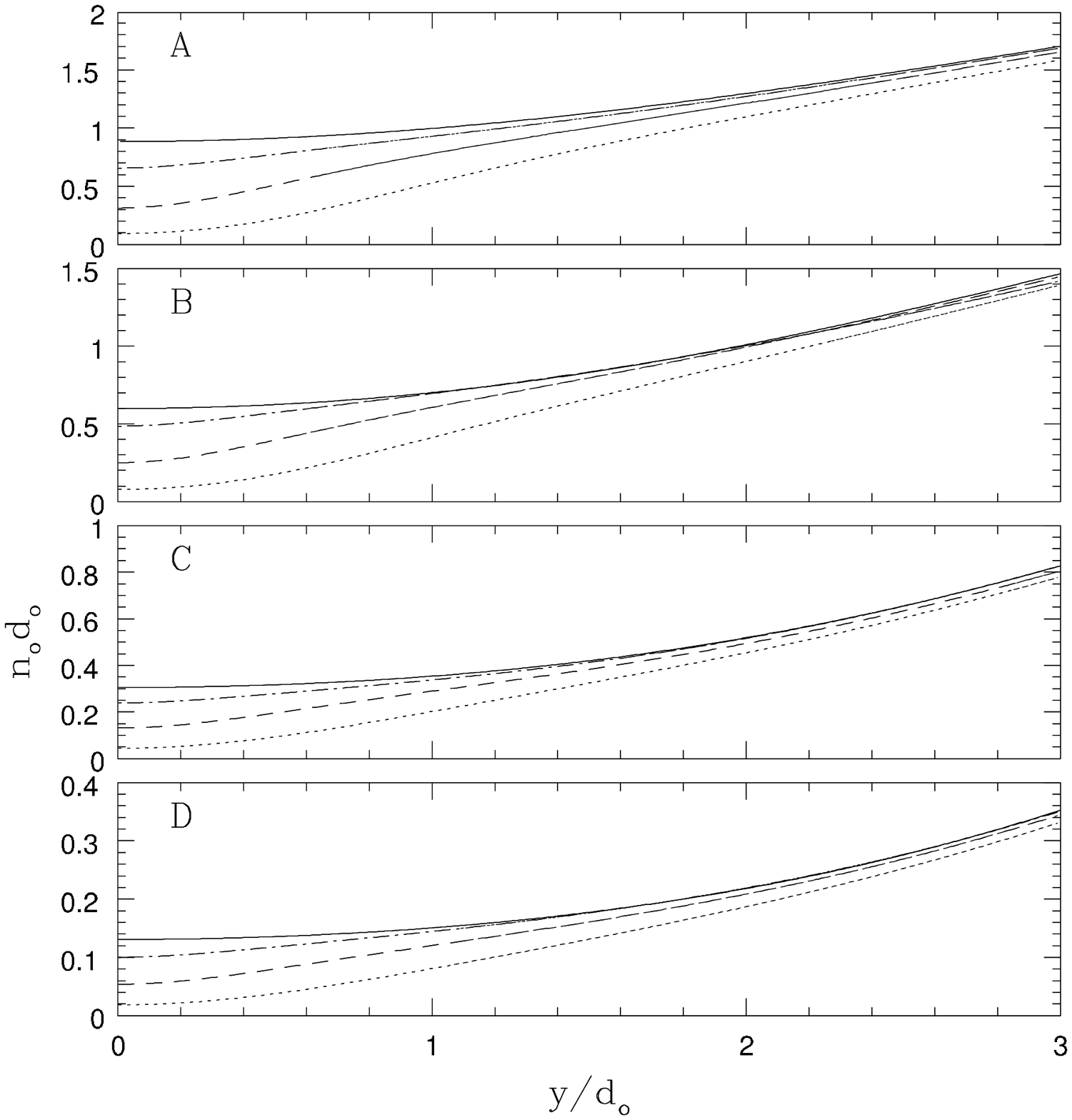}}}
   \caption{The effect of ionization on the surface density for the neutral component ($\Xi\zr=1$). Panels refer to: A to completely dragging case, B C and D respectively to the values 1, 0.3, 0.1 of the parameter $\sigma_{cx}(v_{rel,0})n\zr d\zr$. Solid lines are the solutions without ionization, dotted lines have $ B\zr = 4$, dashed lines have $ B\zr = 1$, dash-dotted lines have $B\zr = 0.25$.}
   \label{fig:6}
\end{figure}
%%%%%%%%%%%%
 %%%%%%%%%%%%
\begin{figure}
\centering{
  
    \resizebox{\hsize}{!}{\includegraphics{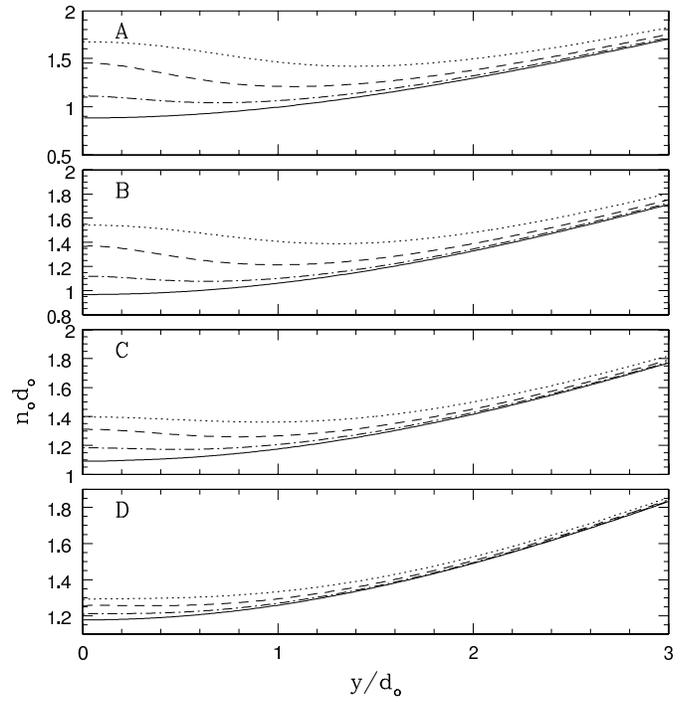}}}
   \caption{The effect of ionization on the surface density for the ionized component ($\Xi\zr=1$). Panels refer to: A to completely dragging case, B C and D respectively to the values 1, 0.3, 0.1 of the parameter $\sigma_{cx}(v_{rel,0})n\zr d\zr$. Solid lines are the solutions without ionization, dotted lines have $ B\zr = 4$, dashed lines have $ B\zr = 1$, dash-dotted lines have $ B\zr = 0.25$.}
   \label{fig:7}
\end{figure}
%%%%%%%%%%%%

\section{\halpha \ luminosity}
Once a neutral component has been introduced in the model, it may be used to evaluate the \halpha \ luminosity of the bow shock nebula and its variation moving away from the head. In Paper I, we have given an approximate estimate of such a luminosity, using some simplifications whose validity were limited to the stagnation point. The knowledge of the global distribution of neutrals and ions in the nebula makes it possible to evaluate the variation of luminosity along the bow shock.

Before investigating the effect of diffusion and charge-exchange we need to discuss the distribution of neutrals in the external layer. In those cases in which the penetration thickness for H atoms if not too high, we assume that the neutrals are homogeneously distributed in the external layer so that the luminosity per unit volume is constant in the direction normal to the layer. But in the highly dragging case it is possible that neutrals cannot reach the contact discontinuity, and remain confined in a thinner shell near the bow shock (al least in the head of the nebula). Anyway only a very high resolution mapping may discriminate among the above cases. In fact the instrumental resolution may spread the pattern. This is a problem if the signal to noise ratio is low so that only the more luminous regions of the nebula are detected.

 %%%%%%%%%%%%
\begin{figure}
\centering{

    \resizebox{\hsize}{!}{\includegraphics{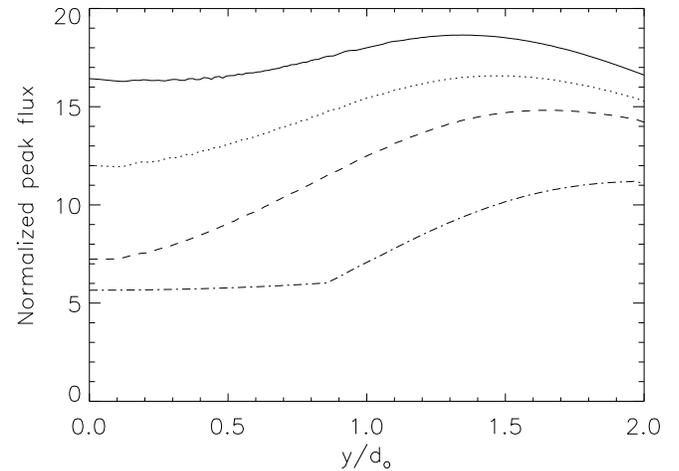}}}
   \caption{Peak luminosity along the nebula from secondary charge-exchange, for the various solution shown in panel A Fig.~5. The flux is in unit of $g_{eff} \,n\zr^{2}\sigma_{cx}(v_{rel,0})v\zr d\zr$ The solid line refers to the completely dragging case without diffusion or ionization; dotted dashed and dot-dashed lines refer respectively to the case $C\zr=0.25, 0.8, 4$}
   \label{fig:8}
\end{figure}
%%%%%%%%%%%%
%%%%%%%%%%%
\begin{figure}
\centering{

    \resizebox{\hsize}{!}{\includegraphics{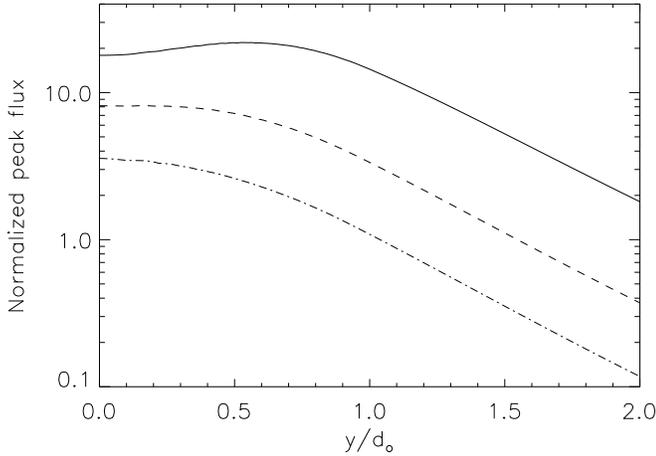}}}
   \caption{Peak luminosity along the nebula from secondary collisional excitation, for the various solution shown in panel A Fig.s~6 7. The flux is in unit of $\epsilon\, n\zr B\zr v\zr$. The solid line refers to the emission in the completely dragging case with an ionization coefficient $B\zr=4$  the dashed line refers to the case $B\zr=1$, and dash-dotted line to the case $B\zr=0.25$ .}
   \label{fig:9}
\end{figure}
%%%%%%%%%%%%

To investigate properly the emission from such nebulae we have to consider the various processes responsible of the excitation of H atoms. They may be divided in two types: primary processes, which involve neutral atoms coming directly from the ambient medium; secondary processes, related to atoms dragged and thermalized in the external layer. In a non-radiative shock moving through a partially neutral medium the dominant processes for optical emission are exciting charge-exchange and collisional excitation of neutral H atoms, which give the observed Balmer spectrum. However primary processes usually have a very low efficiency for incoming neutrals (Paper I) so, the nebula can easily be detected only if the emission is dominated by the secondaries, otherwise a large amount of neutrals in the ISM is required to give enough flux. In what follows only secondary charge-exchange and ionization are considered.

 Fig.~\ref{fig:8} shows the behaviour of peak luminosity along the nebula for secondary charge-exchange. The various curves refer to the same cases of panel A in Fig.~5. There are various effects that contribute to the shape of the \halpha \ luminosity. The solid line represents an upper limit while the others show the effect of diffusion. The luminosity tends to increase as a consequence of the increasing column length of integration, but this effect is reduced by the diminishing of the ionized and neutral density, and this gives the behaviour of the various lines. While the case without diffusion gives a quite constant emission along the nebula, the presence of diffusion reduces the \halpha \ luminosity in the head (where the efficiency of diffusion is grater), and this effect could be used as a tool to determine the physical conditions of such nebulae.

 We have not taken into account the presence of a threshold for the excitation of H atoms: we considered a pulsar so fast with respect to the ambient medium that even in the tail the thermal velocity is above the excitation energy ($\simeq 10$ eV). In the case of a slower pulsar, the temperature in the tail may not be enough to allow exciting charge-exchange, so the \halpha\ flux may be lower. This is the case of PSR J0437+4715 (Bell \& al. \cite{bell95}): only the head of the bow shock nebula have been detected so far despite the fact that this is the nearest object. In fact its velocity is estimated to be $\sim 80$ km/s, very near the threshold, while the other three nebulae, associated with faster pulsars, are detected even in the tail.

Fig.~\ref{fig:9} shows the behaviour of peak luminosity along the nebula for secondary collisional excitation. The various lines refer to the same cases of panel A Fig.~5. Only models with ionization are shown. The \halpha\ emission rapidly decreases in the tail because the rate of interaction is a steep function of the temperature, and vanishes in the tail. So it is possible to discriminate between a flux dominated by charge-exchange or by collisional processes, analyzing the \halpha \ luminosity along the nebula. However even if the efficiency of collisional excitation is one-two orders of magnitude higher than that of charge-exchange, its cross section is much less at typical pulsar velocities, so that only in the faster cases ionization may give a non neglectable contribution, at least in the head of the nebula.

\section{Conclusion}
We have modified the two thin layer solution (Comeron \& Kaper \cite{comeron98}) for bow shock produced by runaway stars (with the aim of deriving a model in the case of pulsar wind bow shock nebulae) with the inclusion of a neutral component. We find that the ISM neutrals lead only to minor effects to the shape of the nebulae (see Fig.~2), but plays an essential role in their luminosity. We have also introduced in our model the possibility that the H atoms, once stopped by the thermalized protons, may diffuse out of the layer or be ionized, as a  consequence of charge-exchange and collisional ionization. Both processes act depleting the neutral surface density in the external layer, but their efficiency is higher in the head of the nebula both as a consequence of the decrease of the rate of charge-exchange and ionization in the tail, and the increase of the geometrical thickness of the layer.

The distributions obtained have been used to evaluate the Balmer emission along the nebula. The study of such emission can be a good tool to understand the properties of local ISM as well as those of the pulsar.  Even if this is not a simple task, and requires very good measurements, which only for two objects are actually available, an accurate study may show which hypothesis can be applied (like, for example, the absence of preionization both from the pulsar surface radiation, or from the emission of shocked ISM). In Paper I, we stressed the point that the total \halpha\ flux depends on the velocity of the pulsar.

 Here we evaluate its variations along the nebula.  The behaviour is different if the dominant process is exciting charge-exchange or collisional excitation, so not only the total flux but also its variation moving away from the stagnation point can be used. Concerning the emission we find that: 1- charge-excange and collisional excitation have different trends; 2- processes like diffusion or ionization may play an essential role changing the emission in the head of the nebula.  This not only allows one to put stronger constraints on the regime of such nebulae, but it may be used to verify the validity of the model, by comparison with the observed cases.

\begin{acknowledgements}
I want to acknowledge Rino Bandiera for the fruitful discussion and the reading of this article. This work has been supported by the Italian Ministry for University and Research (MIUR) under the grant Cofin 2001-02-10.
\end{acknowledgements}

\end{document}